\documentclass[epj+,final]{svjour+}
\usepackage{graphics}

\newcommand{\be}{\begin{equation}}
\newcommand{\bear}{\begin{eqnarray}}
\newcommand{\ee}{\end{equation}}
\newcommand{\ear}{\end{eqnarray}}
\newcommand{\e}{{\rm e}}
\newcommand{\Tr}{{\rm Tr}}
\newcommand{\bel}[1]{\be\label{#1}}
\newcommand{\bearl}[1]{\bear\label{#1}}
\newcommand{\req}[1]{(\ref{#1})}
\renewcommand{\d}{{\rm d}}
\newcommand{\half}{\sfrac12}
\newcommand{\sfrac}[2]{\mbox{$\frac{#1}{#2}$}}
\newcommand{\toto}{\mathop{\longrightarrow}}
\newcommand{\av}[1]{\left\langle{#1}\right\rangle}
\newcommand{\avav}[1]{\left\langle\!\!\left\langle{#1}
           \right\rangle\!\!\right\rangle}
\begin{document}
\title{Time Dependent Local Field Distribution and Metastable States in the SK-Spin-Glass }
\author{Heinz Horner}
\institute{Inst. f. Theoretical Physics, Ruprecht-Karls-University Heidelberg, Germany}

\abstract{
Different sets of metastable states can be reached in glassy systems below some transition temperature depending on initial conditions and details of the dynamics. This is investigated for the Sherrington-Kirkpatrick spin glass model with long ranged interactions. In particular, the time dependent local field distribution and energy are calculated for zero temperature. This is done for a system quenched to zero temperature, slow cooling or simulated annealing, a greedy algorithm  and repeated tapping. Results are obtained from Monte-Carlo simulations and a Master-Fokker-Planck approach. A comparison with replica symmetry broken theory, evaluated in high orders, shows that the energies obtained via dynamics are higher than the ground state energy of replica theory. Tapping and simulated annealing yield on the other hand results which are very close to the ground state energy. The local field distribution tends to zero for small fields. This is in contrast to the Edwards flat measure hypothesis. The distribution of energies obtained for different tapping strengths does again not follow the canonical form proposed by Edwards.
} 
\PACS{ {05.70.Ln} Nonequilibrium and irreversible thermodynamics --  
      45.70.Cc Static sandpiles; granular compaction --
      75.10.Nr Spin-glass and other random models --
      89.75.-k	Complex systems --
      02.60.Pn	Numerical optimization
}
\maketitle
%

\section{Introduction and Summary}
\label{intro}
Complex disordered systems are ubiquitous, in physics and in many other disciplines. Glasses, spin glasses, granular media, structure of proteins, neural networks and various combinatorial optimization problems have a complex organization of low energy states in common. Several methods have been developed over the time in order to deal with the built in disorder. Most widely used is the replica method aiming at the evaluation of the free energy, entropy or at zero temperature the complexity \cite{MPV87}. Typically solutions with spontaneously broken replica symmetry show up below some critical value of temperature or external noise. 

Alternatively some kind of stochastic dynamics has been employed as a tool to investigate such systems \cite{SZ82,Ho92,KT87,CHS93}. For systems with continuous degrees of freedom Langevin dynamics may be used. For systems with discrete degrees of freedom a master equation is appropriate. In particular Glauber dynamics is used for Ising-spins. Dynamical processes of this kind can also be used as algorithm for optimization problems, for example simulated annealing \cite{Ki83}. 

The general picture of complex disordered systems is associated with a rough landscape of energy or free energy with barriers, some of which diverge with the number $N$ of particles or elements sent to $\infty$. Within the approach via dynamics the limit $N\to\infty$ is typically performed first. Only thereafter long time scales are eventually investigated. This means that the barriers diverging with $N$ can not be overcome and the system might be stuck in a certain region of phase space. This means that replica theory, not relying on any kind of dynamics, and the approach via dynamics might lead to different results.

This has consequences for instance in using some kind of dynamical process for finding solutions of combinatorial optimization problems. Replica theory might tell that perfect solutions exist. Dynamics, typically a polynomial algorithm, can indicate that these solutions are not found in polynomial time. But it provides information about suboptimal solutions which can be found in polynomial time. An example, learning in a perceptron with binary couplings, has been discussed \cite{Ho92}.

The dynamical behavior of complex disordered systems on long time scales is crucially affected by the existence of metastable states \cite{Ho92,KT87,CHS93,CS95,Ca04,Ca04a,CL06,As06,Mu06}. Systems undergoing a discontinuous ergodicity breaking transition, e.g. the spherical p-spin-interaction spin glass, show a freezing temperature which is higher than the temperature where single step replica symmetry breaking sets in. This is due to the existence of a large number of meta\-stable states. In systems with continuous ergodicity breaking transition and full replica symmetry breaking, e.g. in the Sherr\-ington-Kirk\-patrick model \cite{SK75}, both temperatures are identical. Nevertheless the states reached at low temperatures for long, but finite, time may be different to those captured by replica theory.

Dynamics in this context has been developed along essentially two different lines. Most investigations are based on two time correlation and response functions of the respective basic degrees of freedom 
\cite{SZ82,Ho92,KT87,CHS93}. Alternatively for spin systems and Glauber dynamics the time dependent local field distribution has been investigated within a combination of a Master- and a Fokker-Planck-equation \cite{Ho90,LCS95,Ea06}. The two methods are in some sense complementary. 

There has been recent interest on metastable states in the context of granular media. Edwards et.al \cite{Ed89,Ed94} have postulated that the steady state reached in granular media after repeated tapping is described by a flat or biased average over all metastable states. The bias is in the form of a Boltzmann factor with some effective temperature characterizing the process of tapping. The number of metastable states has been computed earlier by Edwards and others in an annealed approximation \cite{Ta80,BM80}. This hypothesis was tested with diverse results for various systems and tapping procedures
\cite{BFS02,DL03}. In particular Eastham et.al. \cite{Ea06} have attributed a failure of the Edwards hypothesis to the special distribution of metastable states selected by dynamics.

Metastable states are commonly defined as local minima of a free energy functional obtained for instance by the TAP approach \cite{As06,Mu06}. Investigations of the neighborhood of those minima or other stationary points yield interesting results about the structure of low lying states in different models of disordered systems. Staying completely within the framework of non equilibrium dynamics the concept of a free energy does not apply and one has to rely on different criteria, for instance on the stability of states which are stable with respect to a certain move class. Obviously this is restricted to zero temperature and a state which is stable with respect to one move class might be unstable with respect to a wider move class.

The present contribution resumes the question what kind of metastable states may be reached by various procedures. The analysis is based on the temporal behavior of the local field distribution. In particular the Sherrington-Kirkpatrick  model \cite{SK75} with single spin flip Glauber dynamics is investigated. This is actually a prototype system for a continuous ergodicity breaking transitions or for full replica symmetry breaking. A metastable state in the present context is a state where each spin points in the direction of its local field. This is the field created by the external field and the interaction with other spins. Such a state is stable with respect to single spin flip dynamics. 

Section~\ref{SK-Glauber} contains the definition of the SK-model, the local field distribution and Glauber dynamics, which is also the basis of the Monte-Carlo simulations presented later. The local field distribution obtained from the Edwards measure \cite{Ea06,Ta80,BM80} is discussed in Section~\ref{EdM} and Appendix A. The local field distribution has also been computed in high orders of replica symmetry breaking by Oppermann et.al \cite{OS05}. This is discussed for comparison in Section~\ref{mult-RSB}.
The Master-Fokker-Planck approach is the content of Section~\ref{FP} and Appendix B. Results are presented and discussed for various assumptions about the drift velocity in Sections~\ref{const-v} and \ref{ClosureI}. Results from the closure proposed in \cite{Ho90} are given in Section~\ref{ClosII} and compared with Monte-Carlo simulations in Section~\ref{MC}.

The following results are obtained: 

The local field distribution at zero temperature and late time behaves as $P(k)\sim k$ for $k\to 0^+$. This holds for the following schedules investigated:  Quench from a fully magnetized or random initial state, greedy algorithm, random or thermal tapping and simulated annealing. This behavior is also found within multiple step replica symmetry breaking theory. It contradicts the Edwards flat measure hypothesis. 

The distribution of energies found with tapping follows a displaced Gaussian, where offset and width depend on the tapping strength. According to the Edwards hypothesis the width should be constant and only the offset is supposed to depend on the tapping strength. This is in contrast to the present findings.

Within the framework of a Master-Fokker-Planck approach the drift velocity diverges as $v(k)\sim k^{-1}$ for \linebreak
$k\to 0^+$ and late time. An approximative closure of the resulting hierarchy yields results similar to those obtained from Monte-Carlo simulations at zero temperature.

The ground state energy obtained from replica theory is $E/N\approx-0.763$. Quenching and greedy algorithm yield energies $E/N\approx-0.729$ whereas with repeated tapping or simulated annealing $E/N\approx-0.760$ is found, which is rather close to the ground state energy. The local field distribution is also quite close to the one obtained in replica theory. This indicates that a polynomial algorithm might be able to find good, but suboptimal, solutions to a problem where finding the best solution is a NP-problem. The actual performance certainly depends on on the kind of problem and there might be more efficient polynomial algorithms.

\section{SK-model and Glauber-dynamics}
\label{SK-Glauber}
The energy of the SK-model is given by
\be
H=-\sfrac12 \sum_{ij}J_{ij}\sigma_i\sigma_j-\sum_i h_i\sigma_i.
\ee
where $\sigma_i\!=\!\pm1$ are Ising spins. The couplings $J_{ij}$ are random variables with
\bel{hrsv}
J_{ij}=J_{ji} \qquad\quad \overline{J_{ij}}=0 \qquad \quad
\overline{J_{ij}^{\,2}}= N^{-1}.
\ee
In addition to the external field $h_i$ a spin $\sigma_i$ on site $i$ feels a contribution due to the interaction with other spins. Instead of dealing with the resulting local fields, it is convenient to introduce the product of local field and spin
\bel{nfeq}
k_i=\Big(h_i+\sum_{j (\ne i)} J_{ij} \sigma_j\Big) \sigma_i.
\ee

The central quantity of the present investigation is the distribution of those fields, sorted according to the direction of the spin 
$\sigma_i\!=\!\sigma\!=\!\pm1$
\bel{bwec}
P_\sigma(k)=\sfrac 1N\sum_j \overline{
\av{\delta_{\sigma,\sigma_j}\delta(k-k_j)\Big.}}^J.
\ee
The bar indicates average over the couplings $J_{ij}$ and the brackets denote average over spin configurations. The\linebreak
bonds are assumed to be frozen, but the spin configurations can change in time, resulting in a time dependent distribution $P_\sigma(k;t)$. In general $P_\sigma(k)$ depends on $\sigma$ due to the action of external fields or non symmetric initial conditions.

Unless mentioned otherwise stochastic single spin flip Glauber dynamics is investigated. At a temperature \linebreak
$T\!=\!1/\beta$ the flip rate for a spin at site $i$ is
\bel{mfra}
r(k_i)=\half\Big(1-\tanh(\beta k_i)\Big).
\ee
This is essentially the dynamics used in Monte-Carlo simulations and the time unit corresponds to MC-step per site.

In particular at temperature $T\!=\!0$ the flip rate vanishes for $k>0$ and eventually a stationary state with $k_i\ge0$ for all $i$ is reached. Depending on initial conditions and cooling schedule various metastable states are reached. The distribution of those states depends on initial conditions and cooling schedule.
%
\section{Edwards measure}
\label{EdM}
Tanaka and Edwards \cite{Ta80,BM80} estimated the number of single spin flip metastable states, i.e. states with $k_i\ge0$, in annealed approximation. They find for the SK-model a displaced Gaussian distribution of energies of metastable states
\bel{beap}
P_{MS}(\epsilon)=\sqrt{\frac{N}{2\pi}}\frac{1}{\delta}\,
\e^{-\frac{N}{2\delta^2}(\epsilon-\bar\epsilon)^2}
\ee
where $\epsilon\!=\!E/N$ is the energy per site. They estimated 
$\bar\epsilon\approx 0.5$ and $\delta\!=\!0.31$.

It has been argued \cite{Ed89,Ed94} that a flat or biased average over all metastable states applies to tapped granular systems or spin glasses.  More precisely, the distribution of energies of metastable states obtained with some kind of tapping procedure is supposed to behave as
\bel{bsaq}
P_E(\epsilon)\sim P_{MS}(\epsilon)\,\e^{-\beta_{tap} N \epsilon}
\ee
where the effective temperature $\beta_{tap}$ depends on the\linebreak
strength of tapping. Adopting \req{beap}, $P_E(\epsilon)$ is again a shifted Gaussian with $\bar\epsilon\to\bar\epsilon-\delta^2\,\beta_{tap}$.

The local field distribution in this approximation, in absence of an external field, is a shifted Gaussian truncated at negative values of $k$, see \cite{Ea06} and Appendix A, Eq. \req{evlz}: 
\bel{efvu}
P(k)=\sum_\sigma P_\sigma(k)=\sqrt{\sfrac{2}{\pi}}\,
\frac{\e^{-\frac12 (k-\kappa)^2}}
{1+{\rm erf}\Big(\frac{1}{\sqrt{2}} \kappa\Big)} \,\Theta(k)
\ee
with 
\bel{enua}
\kappa=\sfrac12\{P(0^+)+\beta_{tap}\}.
\ee
This distribution has a discontinuity of size $P(0^+)$ at \mbox{$k\!=\!0$}.
The energy per site is
$\epsilon\!=\!-\frac12[ P(0^+)+\kappa]$.

\begin{figure}
\center{ 
\resizebox{0.47\textwidth}{!}{
\includegraphics{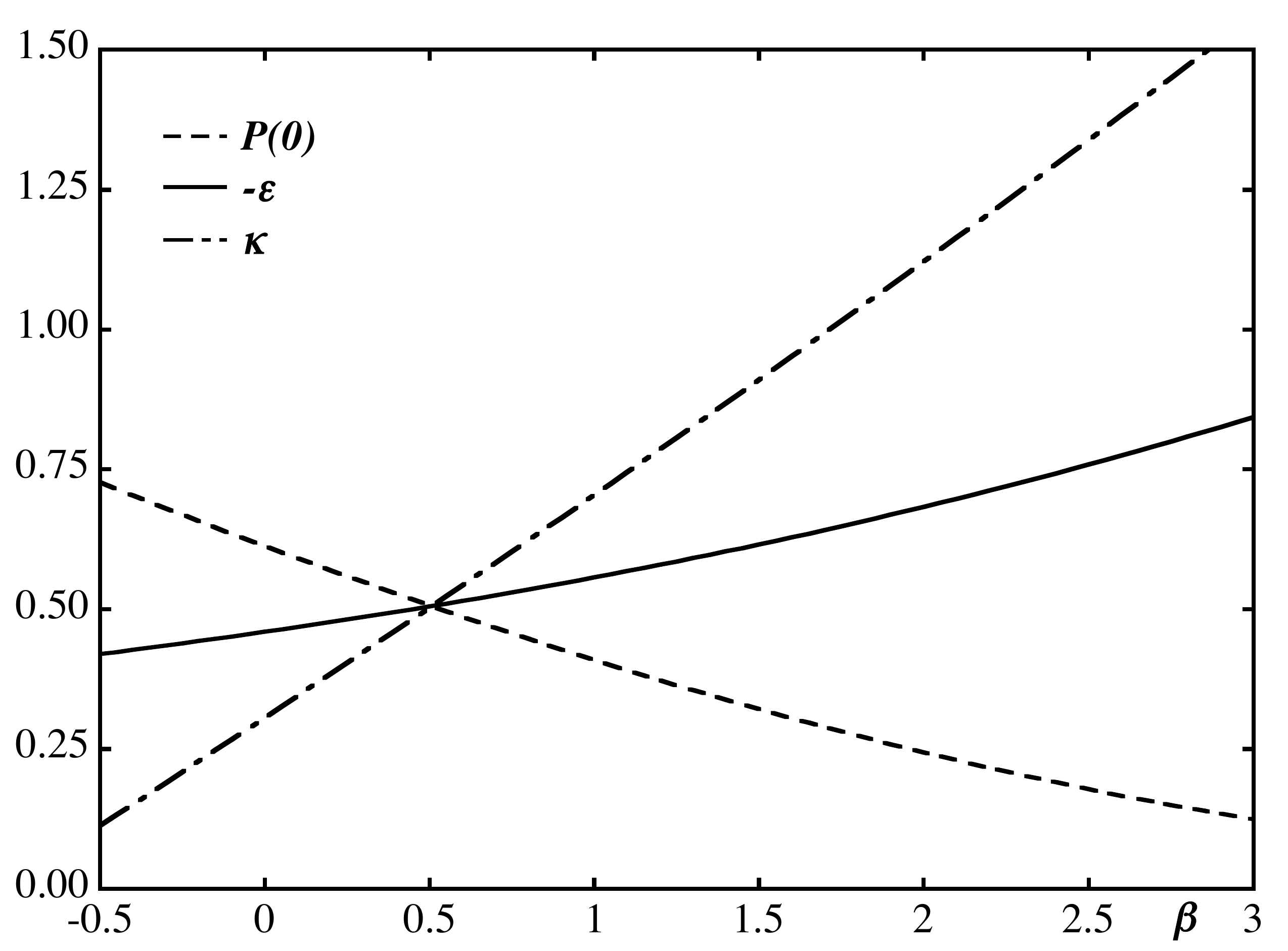}}}%
\caption{Energy $\epsilon$ per site \req{ljtx}, local field distribution 
$P(0^+)$ and position $\kappa$ of the maximum of $P(k)$ as function of the effective inverse temperature $\beta_{tap}$ calculated from the Edwards measure.}
\label{flatav} 
\end{figure}

\begin{figure}
\center{ 
\resizebox{0.462\textwidth}{!}{
\includegraphics{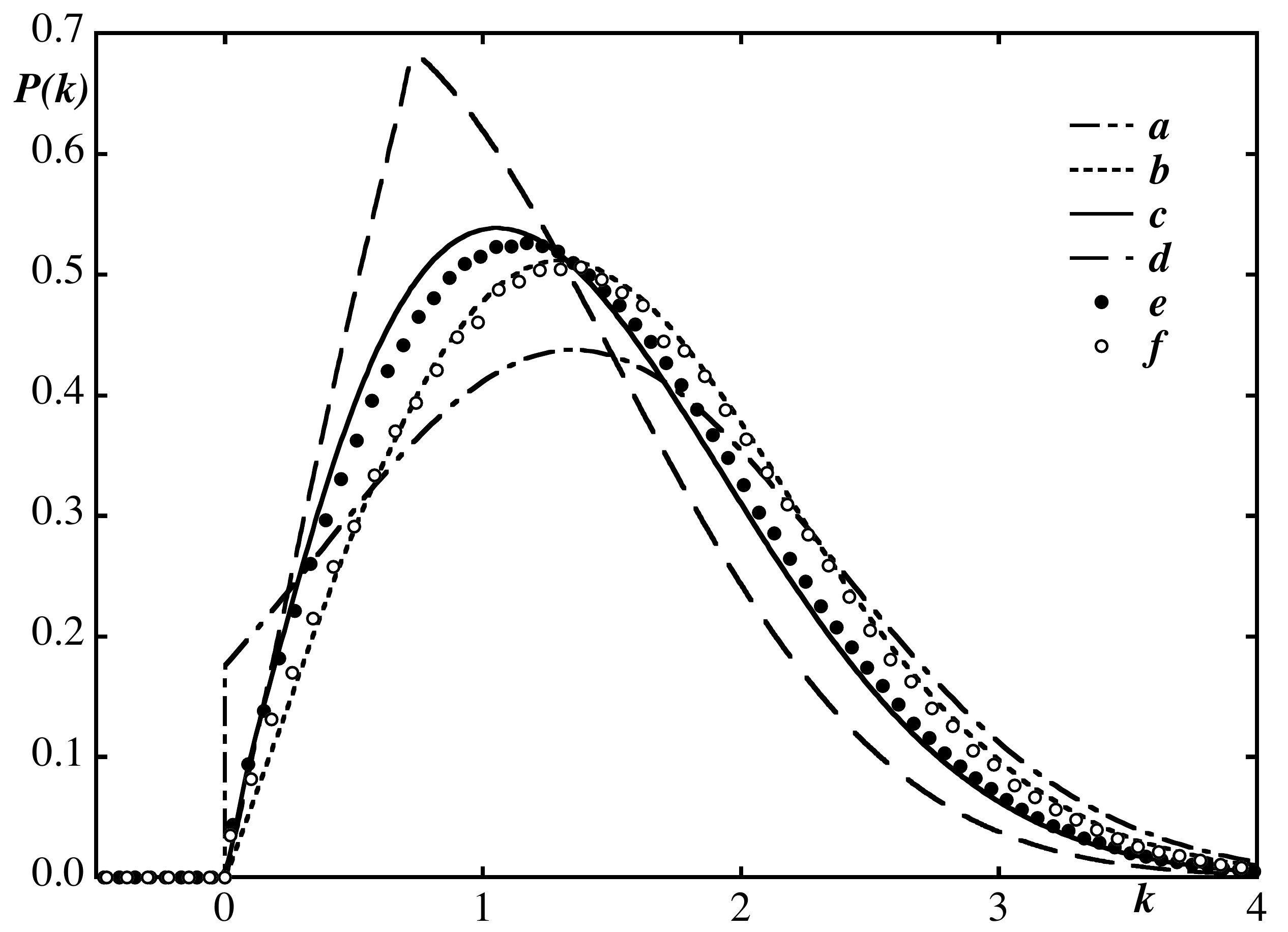}}} %
\caption{Local field distributions: $a$) Edwards measure, Eq.~\req{efvu}; 
$b$) multi step RSB \cite{OS05}; $c$) Master-Fokker-Planck equation; 
$d$)~simplified Master-Fokker-Planck equation; $e$) Monte Carlo simulation at zero temperature; $f$) Monte Carlo simulation with random tapping.} 
\label{P(k)}
\end{figure}

Eqs. \req{efvu} and \req{enua} can be solved numerically for \mbox{$k\!=\!0^+$} 
as function of the bias $\beta_{tap}$. The resulting energy 
$\epsilon$, discontinuity of the local field distribution $P(0^+)$ and peak position $\kappa$ are shown in Fig.\,\ref{flatav}. In Fig.\,\ref{P(k)} the local field distribution \req{efvu} for $\beta\!=\!2.5$ is plotted together with other results  discussed later. For the above bias $\epsilon\!=\!-0.76$, 
$\kappa\!=\!1.35$ and 
$P(0^+)\!=\!0.175$. The bias is choosen such that the energy $\epsilon$ is close to the ground state energy obtained in replica theory \cite{OS05}.
%
%
%
%
\section{Multi step replica symmetry breaking}
\label{mult-RSB}
A linear behavior with slope $0.3$ of the local field distribution for small fields at zero temperature has been found by Sommers and Dupond \cite{SD84}.
Recently Oppermann and coworkers \cite{OS05} have investigated the replica symmetry breaking solution for $T\!=\!0$ in high order. They find 
$\epsilon\!=\!-0.763\cdots$ and again a linear behavior of the local field distribution with slope $0.3$. The same values have been obtained by Pankov \cite{Pa06}. The present investigation deals with the distribution of 
the product of local field and spin $k_i$, Eq.~\req{nfeq}, resulting in 
$P(k)\!\to\! 0.6\,k\,$ for $k\!\to\! 0^+$ adopting the above value. The complete form of the local field distribution is also shown in 
Fig.\,\ref{P(k)}.

Even for a bias chosen such that the energy agrees for both approaches, $P(k)$ is quite different, indicating that the replica calculation and the Edwards measure refer to different states. This is actually expected since the replica calculation is supposed to present an average over true ground states, whereas the Edwards measure is supposed to be a biased average over all metastable states.

%
\section{Master-Fokker-Planck equation}
\label{FP}
An equation of motion for the local field distribution has been derived in 
\cite{Ho90}. The following contains a slightly simplified version.

A spin flip $\sigma_i\to-\sigma_i$ at site $i$ results in changes of the modified local fields
\bel{dqvi}
k_i\to -k_i
\ee
and for $j\ne i$
\bel{kwnr}
k_j\to k_j-2J_{ij}\sigma_i\sigma_j.
\ee
With a flip rate $r(k)$, Eq. \req{mfra}, the resulting time dependence of the local field distribution is
\bearl{kzsc}
\!\!&&\!\partial_t P_\sigma(k,t)=r(-k)P_{-\sigma}(-k)-r(k)P_\sigma(k)+\bigg.
\\
\!\!&&+\sfrac1N \!\sum_{i\ne j}\!\overline{
\av{\!r(k_j)\delta_{\sigma,\sigma_i}\!
\Big\{\delta(k-k_i+2\sigma_iJ_{ij}\sigma_j)-\delta(k-k_i)\Big\}\bigg.\!}}^J
\nonumber
\ear
where the first two terms are due to \req{dqvi} and the last term is due to \req{kwnr}. Performing the average over the couplings, neglecting contributions of order $N^{-1}$, introducing a "diffusion constant" (see later)
\bel{kjhw}
D(t)=2\avav{r\big.(t)}=2\sum_{\sigma}\!\int\!\d k\,r(k)P_\sigma(k,t)
\ee
and a two point function
\bearl{kury}
&&R_{\sigma \sigma'}(k,k')=
\nonumber\\
&& \quad \sfrac1N\sum_{i\ne j}
\overline{\av{\delta_{\sigma,\sigma_i}\delta(k-k_i)
\sigma_iJ_{ij}\sigma_j\Big.\delta_{\sigma',\sigma_j}\delta(k'-k_j)\bigg.}}^J
\quad
\ear
Eq. \req{kzsc} is written as
\bearl{thjs}
\partial_t P_\sigma(k,t)&=&r(-k)P_{-\sigma}(-k,t)-r(k)P_\sigma(k,t)
\nonumber\\
&+&2\partial_k \sum_{\sigma'}\!\int\!\d k'\,R_{\sigma\sigma'}(k,k';t)r(k')
\nonumber\\
&+&D(t) \partial_k^2 P_\sigma(k,t).
\ear
The double bracket $\avav{\cdots\big.}$ indicates average over $P_\sigma(k,t)$.
Introducing a drift velocity $v_\sigma(k)$
\bel{ftho}
v_\sigma(k)P_\sigma(k)=-2\sum_{\sigma'}\!\int\!\d k'\,R_{\sigma \sigma'}(k,k')
r(k')
\ee
Eq. \req{thjs} combines elements of a Master-equation and a Fokker-Planck equation:
\bearl{dast}
\partial_t P_\sigma(k,t)&=&
r(-k)P_{-\sigma}(-k,t)-r(k)P_\sigma(k,t)
\nonumber\\
&-&\partial_k \Big(v_\sigma(k,t) -D(t)\partial_k\Big) P_\sigma(k,t).\;\; \quad
\ear
The two point function \req{kury} obeys 
\bel{szhw}
kP_\sigma(k)=\sum_{\sigma'}\!\int\!\d k'\,R_{\sigma \sigma'}(k,k')
\ee
and with \req{ftho} the following sum rule is obtained
\bel{nsza}
\avav{v\big.}=-2\avav{kr\big.}.
\ee

Along the same way an equation of motion can be derived for the two point function \req{kury}. The derivation is sketched in Appendix B. It involves, however, a three point function, and ultimately a hierarchy of equations is generated which requires some kind of truncation.

The probability to flip a spin in field $k$ per unit time is $r(k)$. Let 
$\tau(t)$ be the probability that a given spin has flipped within time $t$. 
Then
\bel{garc}
\partial_t \tau(t)=\sum_{\sigma}\!\int\!\d k\,r(k)\,P_\sigma(k,t)
=\half D(t).
\ee
This quantity has actually been used in \cite{Ea06} as a measure of time.
%
\subsection{Initial state}
\label{ini-state}
For any factorizing initial state which is not correlated with the couplings 
$J_{ij}$, the average over $J$ in \req{kury} involves only the $J$-dependence of the fields $k_i$, Eq. \req{nfeq}, resulting in
\bearl{clis}
&&\!\!\!R_{\sigma \sigma'}(k,k')
\nonumber\\
&& \ \approx -\Big(\partial_k+\partial_{k'}\Big)
\sfrac1{N^2}\!\sum_{i\ne j}
\overline{\av{\!\delta_{\sigma,\sigma_i}\delta(k-k_i)
\Big.\delta_{\sigma',\sigma_j}\delta(k'-k_j)\bigg.\!}}^J
\nonumber\\
&& \ =-\Big(\partial_k+\partial_{k'}\Big) P_\sigma(k)P_{\sigma'}(k').
\ear
Examples are a fully magnetized initial state with $\sigma_i\!=\!1$ for all $i$, or a state with random spins. 
In absence of an external field a fully magnetized initial state is actually equivalent to any other initial spin configuration within the ensemble of couplings. The magnetization plays then the role of the overlap with this initial state.

In the following a fully magnetized initial state is assumed. The initial value of the local field distribution is
\be
P_{+1}(k,0)=\sfrac{1}{\sqrt{2\pi}}\,\e^{-\frac12 k^2} \qquad
P_{-1}(k,0)=0.
\ee
The diffusion constant, Eq. \req{kjhw}, is $D(0)\!=\!1$ and the drift velocity obtained from \req{ftho} and \req{clis} is
\bel{rbpq}
v_\sigma(k,0)=-2\avav{kr\big.(0)}-k.
\ee
%
\subsection{Constant and linear drift velocity}
\label{const-v}
Investigating the validity or failure of the Edwards hypothesis \cite{Ed94,Ed89}
Eastham et al. \cite{Ea06} have assumed a constant drift velocity 
$v(k,t)\!=\!c\,D(t)$ and have solved \mbox{Eq.~\req{dast}} numerically for $T\!=\!0$ and time up to $t\sim 10$. At $t\!=\!10$ they find reasonable over all agreement with MC-simula\-tions. There appears, however, a slightly smeared out step at $k\!=\!0$.

The asymptotic behavior at late time can easily be evaluated for $v(k)\!=\!(c-c'k)\,D(t)$. Neglecting for a moment the non local contributions in \req{dast}, a Gaussian centered at $k\!=\!\kappa(t)$ and with width $\Delta_k(t)$ results
\bel{luhw}
P(k,t)\sim \e^{-(k-\kappa(t))^2/2\Delta_k^2(t)}.
\ee
Introducing $\tau(t)\!=\!\int_0^t D(t)\d t$
\be
\kappa(t)=\frac{c}{c'}\Big(1-\e^{- c'\tau}\Big)
\ee
the width is
\be
\Delta_k^2(t)=\frac{1}{c'}\Big(1-(1-c')\,\e^{-2c'\tau}\Big).
\ee

For $T\!=\!0$ and late time the part of the local field distribution extending to 
$k<0$ is expected to be small. This means that $D(t)$, Eq. \req{kjhw}, is small. 
For $k<0$ the dominant contributions to \req{dast} are the second term and the term involving the second derivative, i.e.
\be
-P(k,t)+D(t)\partial_k^2\,P(k,t)\approx0 \qquad\mbox{for} \;\; k<0
\ee
which is solved by
\be
P(k,t)=a(t)\,\e^{k/\sqrt{D(t)}},
\ee
and with \req{kjhw} \ $D(t)\!=\!2a^2(t)$. Similar arguments lead to
\be
P(k,t)=a(t)\Big(2-\e^{-k/\sqrt{D(t)}}\Big)
\ee
for small $k>0$. The factor $a(t)$ is obtained by matching with the solution \req{luhw} at $k\approx 0$, i.e. 
\be
a(t)\approx \frac{1}{2\sqrt{2\pi\Delta(t)}}\e^{-\kappa^2(t)/2\Delta^2(t)}.
\ee
Especially for $t\to\infty$ one obtains
\be
\kappa\to\frac{c}{c'} \qquad \Delta \to \frac{1}{\sqrt{c'}} \qquad
D\to\frac{c'{}^2}{8\pi}\,\e^{-c^2/c'}.
\ee

This solution resembles the in some sense the distribution obtained from the Edwards measure \req{efvu} with $\beta\!=\!2c$ and $c'\!=\!1$. The step at $k\!=\!0$ is, however, smeared out and even at late time there is a tail of $P(k)$ extending to $k<0$ indicating that the above assumption regarding the drift velocity is not appropriate for metastable states with $P(k<0)\!=\!0$.

For $c'\to 0$ and $t\to\infty$ the step at $k\!=\!0$ vanishes, but $\kappa$ and with it the energy per site grow without limit. The qualitative agreement with the MC-data in \cite{Ea06} appears to be a consequence of the particular choice of $c$ and an appropriate finite $t$ in this investigation. It has been argued that the results at finite time are actually more appropriate for the asymptotic behavior of a system of finite size. I shall come back to this point later.
%
\subsection{Qualitative discussion of the drift term at T=0}
\label{Quali}
At zero temperature and for late time metastable states are reached. this means that all local fields are expected to be non negative, i.e. 
$P(k<0,t\to\infty)\to 0$.
Assume that for small $k$ and late time
\be
P_\sigma(k,t)\to c\,k^\alpha \,\Theta(k).
\ee
This gives rise to a contribution $\sim k^{\alpha-2}$ in the diffusion term of Master-Fokker-Planck equation \req{dast}, which has to be compensated by the drift term. This results in
\be
v_\sigma(k,t)\to \alpha\,D(t)\,k^{-1}.
\ee
The evaluation of the actual value of $\alpha$ requires to solve the equations for the $n$-point functions of higher order.
%
\subsection{Closure I: Disregarding dynamical correlations}
\label{ClosureI}
The two point function \req{kury} involves an average over the couplings 
$J_{ij}$. The couplings are contained in the local fields $k_i$, \req{nfeq}. In addition correlations between the couplings and the actual states reached at finite $t$ build up in the course of time. Neglecting those  Eq. \req{clis} holds for all times. The resulting drift velocity is
\bel{wnoa}
v_\sigma(k,t)=D(t)\partial_k \ln\Big(P_\sigma(k)\Big)-2\avav{r'(t)\big.}
\ee
and the equation of motion \req{dast} becomes
\bearl{drfn}
\partial_t P_\sigma(k,t) &\approx& r(-k)P_{-\sigma}(-k,t)-r(k)P_\sigma(k,t)
\nonumber\\
&+&2 \avav{r'(t)\big.} \partial_k P_\sigma(k,t).
\ear

The qualitative behavior for $T\!=\!0$ is easily discussed. For small $k$ the local field distribution is 
\be
P_\sigma(k,t)\approx P_\sigma(0,t)+k\,P'_\sigma(0^{\pm},t)
\ee
with $P_+(0,t)\ne P_-(0,t)$ and $P'_\sigma(0^{+},t)\ne P'_\sigma(0^{-},t)$.
For $k\to 0$ Eq. \req{drfn} yields
\bear
&&\partial_t \sum_\sigma P_\sigma(0,t)
\\
&& \qquad =-2 \sum_\sigma P_\sigma(0,t)
\sum_{\sigma'}\Big(P'_{\sigma'}(0^+,t)-P'_{\sigma'}(0^-,t)\Big)
\nonumber
\ear
which is solved by $\sum_\sigma P_\sigma(0,t)\to 0$ for $t\to\infty$ with finite $P'_\sigma(0^+,t)>0$. The drift velocity obeys for $k>0$
\be
v_\sigma(k,t)\toto_{k\to 0} D(t) \frac{P'_\sigma(0^+,t)}
{P_\sigma(0,t)+k\,P'_\sigma(0^+,t)}
Ñ\toto_{t\to\infty} \frac{D(t)}{k}
\ee
and the exponent $\alpha$ introduced in the previous subsection is 
$\alpha\!=\!1$.

This is supported by the numerical integration of\linebreak Eq.~\req{drfn}. The complete local field distribution for $t\!\to\!\infty$ is also shown in Fig.\,\ref{P(k)}. 
The slope of $P(k)$ for $k\to0$ is now $0.9$. This increased value is in accordance with the maximum of $P(k)$ shifted towards smaller values of $k$ and a higher value of the energy $\epsilon$. This behavior is also found in the investigations reported later.

Already this rather simple closure shows that the assumption of a constant or linear drift velocity, \cite{Ea06} and Sect.\,\ref{const-v}, is not appropriate. It also shows that the Edwards measure \cite{Ta80} does not apply to the situation captured by the Master-Fokker-Planck equation \req{dast}. It has to be stressed again, that this equation describes a rapid quench from a fully magnetized or high temperature state to $T\!=\!0$, and that the limit 
$N\!\to\!\infty$ is performed first. In terms of combinatorial optimization problems this corresponds to a greedy search and therefore to a polynomial algorithm $\sim \! N^2$. An average over all metastable states, on the other hand, would require an exponential effort.
%
\subsection{Closure II: Approximative treatment of dynamical correlations}
\label{ClosII}
An improved theory taking into account dynamical correlations has been proposed by the author \cite{Ho90}. It is based on a  factorization of the three point function entering the equation of motion of the two point function \req{kury}. It involves a modified Kirkwood superposition approximation known from the theory of real gasses \cite {Ki35}. A slightly simplified version is outlined in Appendix B. The Master-Fokker-Planck equation \req{dast} and the equation for the drift velocity \req{hatb} can be integrated numerically with the initial conditions discussed in Sect.\,\ref{ini-state}. For technical reasons the calculations are performed for a small finite temperature, typically $T\!=\!0.01$, much smaller than the freezing temperature $T_c\!=\!1$.

\begin{figure}
\center{ 
\resizebox{0.47\textwidth}{!}{
\includegraphics{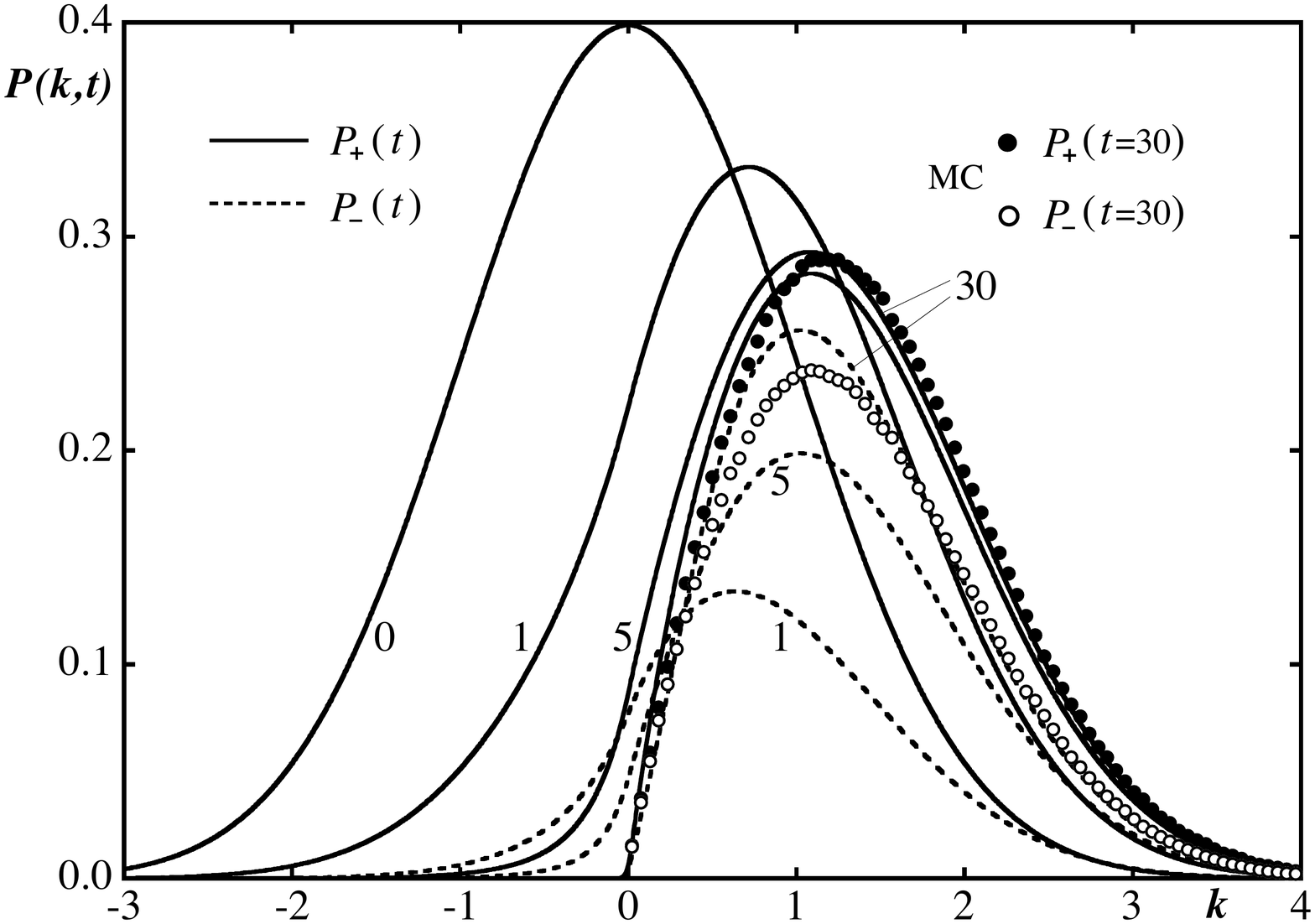}}} %
\caption{Local field distribution $P_\sigma(k,t)$ for $t\!=\!0$, $1$, $5$ and $30$ obtained from the Master-Fokker-Planck approach and from Monte-Carlo simulations for $t\!=\!30$.}
\label{P(k,t)} 
\end{figure}
\begin{figure}
\center{ 
\resizebox{0.47\textwidth}{!}{
\includegraphics{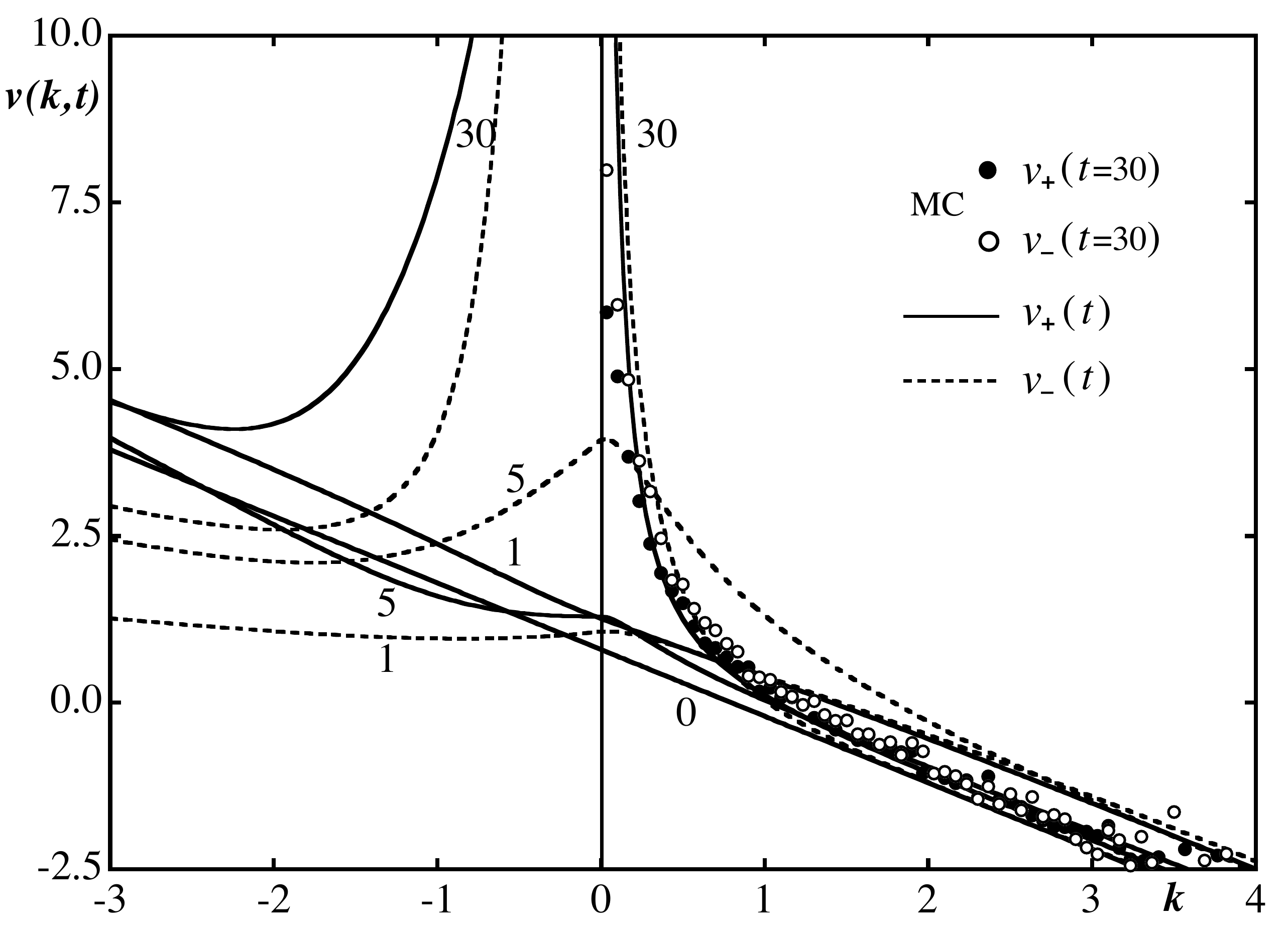}}} %
\caption{Drift velocity $v_\sigma(k,t)$  for $t\!=\!0$, $1$, $5$ and $30$ obtained from the Master-Fokker-Planck approach and  from Monte-Carlo simulations for $t\!=\!30$.}
\label{v(k,t)} 
\end{figure}
%
Fig.\,\ref{P(k,t)} shows the resulting local field distributions for spin up and down, respectively, for times $0$, $1$, $5$ and $30$. The result from the Monte-Carlo simulation discussed in the next section for $t\!=\!30$ are also shown. The complete local field distribution $P(k)\!=\!P_+(k)+P_-(k)$ is compared with other results in Fig.\,\ref{P(k)}. This shows that for long time $P(k,t)\to 0$ for $k\to 0$ in qualitative agreement with replica theory and in contrast to the finite step obtained from Edwards hypothesis. The value of the exponent $\alpha$ introduced in Sect. \ref{Quali} is consistent with $\alpha\!=\!1$.

The agreement between Monte-Carlo simulation and the Master-Fokker-Planck approach with the present closure is quite satisfactory. This indicates that the present closure captures the essential mechanisms: the effect of flipping a spin in its own local field, described by the first two terms in \req{dast}, and the small effects on the local fields of all the other spins, described by the drift- and diffusion terms of the Master-Fokker-Planck equation \req{dast}.

The drift velocity for spin up and down for the same times as above is plotted in Fig.\,\ref{v(k,t)} together with the results of the simulation at the latest time. The formation of the $1/k$-divergence of $v_\sigma(k,t)$ for long time and $k\to 0$ is seen in the Master-Fokker-Planck approach as well as in the Monte-Carlo data.
%
\begin{figure}
\center{ 
\resizebox{0.46\textwidth}{!}{
\includegraphics{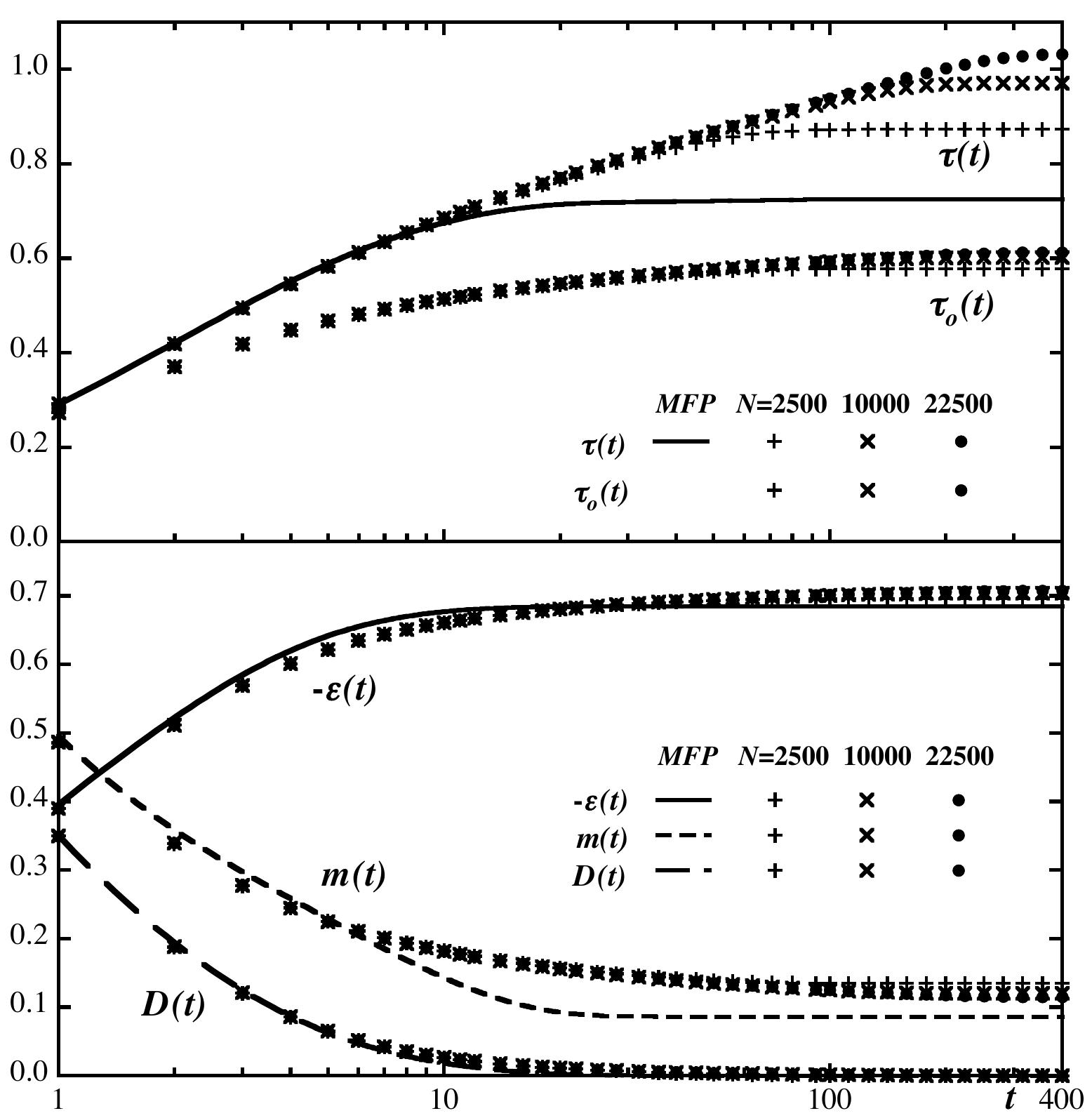}}} %
\caption{Energy per site $\epsilon(t)$, magnetization $m(t)$, diffusion 
constant $D(t)$, flips per site $\tau(t)$ and fraction $\tau_0(t)$ of spins fliped at least once obtained from the Master-Fokker-Planck approach and from zero temperature Monte-Carlo simulations. The time unit corresponds to Monte-Carlo steps per site.}
\label{F(t)} 
\end{figure}

Energy per site $\epsilon(t)$, magnetization $m(t)$, diffusion constant 
$D(t)$ and spin flips per site $\tau(t)$, see Eq. \req{garc}, are plotted as functions of time in Fig.\,\ref{F(t)}. Again Master-Fokker-Planck and Monte-Carlo results are compared. Asymptotic values for $t\to\infty$ are listed in 
Table \ref{Table}. They are discussed later.

%
\section{Monte-Carlo simulations}
\label{MC}
In order to test the results of the previous section and in order to investigate alternative optimization methods various Monte-Carlo simulations have been performed. The general procedure is standard. A site $i$ is selected at random and flipped with a probability $r(k_i)$, Eq. \req{mfra}. The local fields are updated according to \req{kwnr}
\be
k_i\to -k_i \qquad \quad k_j\to k_j-2J_{ij}\sigma_i \sigma_j.
\ee
The local field distribution, Eq. \req{bwec}, two point function, 
Eq. \req{kury} and drift velocity, Eq. \req{ftho}, are also updated accordingly.

Alternatively a greedy algorithm is investigated, where in each step the spin  
$\sigma_i$ withe most negative value of $k_i$ is selected and flipped, until $k_i\ge 0$ for all $i$.

For the SK-model in the limit $N\to\infty$ only the first and second moment of the distributions of couplings \req{hrsv} is relevant. The simulations are therefore performed with $J_{ij}\!=\!\pm 1/\sqrt{N}$ for $i\ne j$ and $J_{ii}\!=\!0$. This speeds up the computation considerably and the local fields are integer  multiples of $1/\sqrt{N}$.

The simulations are performed on samples with up to $N=22500$ sites. The results are typically averaged over $1000$ runs with $200$ different sets of random couplings $J_{ij}$.
%
\subsection{Zero temperature simulation and greedy algorithm}
\label{MC-T0}
Results of the zero temperature Monte-Carlo simulations and from the greedy algorithm are shown in 
Figs.\,\ref{P(k)} to \ref{F(t)} and in Table \ref{Table}.

Fig.\,\ref{F(t)} compares the time dependence of various quantities obtained from zero temperature Monte Carlo simulations on samples of different size $N$ and from the Master-Fokker-Planck approach. Regarding the energy, magnetization and diffusion constant there are some discrepancies at intermediate times, the behavior at early and late time is, however, reasonably well reproduced.
 
Energy and magnetization can be fitted according to
\bel{vota}
F(N)\approx F_0+F'\,N^{-\alpha}.
\ee
In \cite{Pa03} $\alpha=0.33$ was proposed. An improved fit to the present data is obtained with $\alpha=0.2$. This yields the values shown in 
Tab.\,\ref{N-fit}. The asymptotic values of the energy are slightly lower than those obtained in \cite{Pa03}. This is due to the different choice of 
$\alpha$ and the fact that larger values of $N$ are used in the present investigation.

Starting with a fully magnetized initial state the magnetization at $T\!=\!0$ remains finite. This is also true for the greedy algorithm. The values obtained by the simulations are in reasonable agreement with the result from the Master-Fokker-Planck approach.

The local field distribution $P(k)$ obtained from the greedy algorithm and from zero temperature Monte-Carlo simulations are almost identical. For finite $N$ 
there is a small step at $k=0$ vanishing as
\bel{vead}
P(k=0^+,N)\approx 1.4\,N^{-0.46}
\ee
which is in reasonable agreement with the result obtained in \cite{Pa03}.

In the limit $N\to\infty$ \ $P_{\pm}(k) \sim k$ for $k\to 0$ is found in all three cases. This is in contrast to the step resulting from the Edwards hypothesis.

The distribution of the energy per site $P_E(\epsilon)$, see \req{beap}, is fitted well by a shifted Gaussian, with $\bar \epsilon \approx-0.70$ and $\delta\approx 0.62$ for $N=10000$. Similar values are found for other values of $N$.

\begin{table}[top]
\center{
\begin{tabular}{lll}
\hline\noalign{\smallskip}
  & \  $F_0$ & \ $F'$ \\
\noalign{\smallskip}\hline\noalign{\smallskip}
$\epsilon_{\rm MC}$ & -0.729 & 0.157 \\
$\epsilon_{\rm greedy}$ & -0.730 & 0.182 \\
$m_{\rm MC}$ & 0.074 & 0.29 \\
$m_{\rm greedy}$ & 0.091 & 0.53 \\
\noalign{\smallskip}\hline
\end{tabular}}
\caption{Coefficient of fitting energy per site $-\epsilon(N)$ and magnetization 
$m(N)$ according to \req{vota} with $\alpha=0.2$.}
\label{N-fit} 
\end{table}

A significant discrepancy between the results of the Master-Fokker-Planck approach and Monte Carlo simulations shows up for the number of spin flips measured as $\tau(t)$. A given spin might actually flip more than once. Let $\tau_0(t)$ be the fraction of spins which has flipped at least once.Then the fraction of spins which have not flipped at all is $1-\tau_0(t)$. This quantity is also shown in Fig.\,\ref{F(t)}. The difference between $\tau(t)$ and 
$\tau_0(t)$ is a measure of multiple spin flips. The results obtained for samples of different size follow universal curves up to some $N$-dependent time 
$\bar t(N)$. A fit covering the range $t>10$ yields
\be
\bar t(N)\approx 0.22\,N^{0.7}
\ee
and for $\tau(t,N)$ and $\tau_0(t,N)$, respectively, the asymptotic values
\bearl{bsel}
\tau(\infty,N)&\approx& 0.26+0.077\,\ln(N)
\nonumber\\
\tau(t,\infty)&\approx& 0.43+0.11\,\ln(t)
\nonumber\\
\tau_0(\infty,N)&\approx& 0.66-0.19\,N^{-0.25}
\nonumber\\
\tau_0(t,\infty)&\approx& 0.66-0.32\,t^{-0.36}
\ear
are obtained. This means that $\tau(t,n)$ diverges logarithmically for 
$t\to\infty$ and $N\to\infty$ whereas $\tau_0(t,N)$ remains finite in this limit. A divergence of $\tau(t,N)$ was also found by Eastham et.al. \cite{Ea06} on the basis of a constant drift velocity. A fit corresponding to Eq. \req{bsel} would, however, yield rather different values. The Master-Fokker-Planck approach yields a finite asymptotic value for \mbox{$\tau(t\!\to\!\infty)$} which is not too far from the asymptotic value of $\tau_0(\infty,\infty)$.
%
\subsection{Tapping Dynamics}
\label{MC-tap}

An investigation of tapping dynamics is of interest not only in the context of the Edwards hypothesis \cite{Ed89}. Tapping or thermal cycling might also be used for combinatorial optimization problems \cite{Hof05}. Tapping might be done either by heating the system periodically to some temperature $T_{tap}$ or by flipping randomly selected spins with probability $p$. 

The minimal energy obtained within $n_{tap}$ cycles of random tapping is fitted to
\be
\epsilon_{tap}(n_{tap},p,N)=\epsilon_{0,tap}(p,N)
-\epsilon'_{tap}(p,N)\,n_{tap}^{-\alpha'}
\ee
with $\alpha'=0.5$. There is no significant $N$-dependence in 
$\epsilon_{0,tap}(p,N)$ and the optimal value 
$\epsilon_{0,tap}(p=0.1)\approx -0.759$ is obtained for $p=0.1$.

Thermal tapping yields similar results with an optimal tapping temperature 
$T_{tap}\approx 0.7$. Tapping is obviously quite effective in finding states with low energy. 

\begin{table}
\center{
\begin{tabular}{llll}
\hline\noalign{\smallskip}
Method & \  $-\epsilon$ & \ $m$ \\
\noalign{\smallskip}\hline\noalign{\smallskip}
Multi step RSB & 0.763 \\
Eswards measure $\beta\!=\!2.5$ & 0.760 \\
Master-Fokker-Planck eq. & 0.688 & 0.086  \\
Monte Carlo \ $T\!=\!0$ & 0.729 & 0.074  \\
Greedy  & 0.730 & 0.091  \\
Random tapping, $p\!=\!10$\%  & 0.760 \\
Thermal tapping, $T_{tap}\!=\!0.7$  & 0.759 \\Simulated annealing & 0.759 \\
\noalign{\smallskip}\hline
\end{tabular}}
\caption{Energy per site $\epsilon$ and magnetization $m$ obtained from the various methods discussed in the text. For the simulations extrapolated values for $N\to\infty$ are shown.}
\label{Table} 
\end{table}

For comparison simulated annealing has also been\linebreak
 tested. A schedule 
$T(t)\!=\!(1-t/\bar t)\,T_0$ with $T_0\!=\!1.5$ and $\bar t\!=\!300\cdots 10000$ has been used. For the slowest cooling schedule $\epsilon\approx -0.758$ has been found.

The local field distribution obtained with the tapping procedure is shown in 
Fig.\,\ref{P(k)}. The agreement with the local field distribution obtained by multi step RSB is quite good. Again there is a small step of $P(k,N)$ at $k=0$, which vanishes for $N\to\infty$ similar to Eq. \req{vead} with a reduced prefactor $\sim 0.5$. Even at finite $N$ \ $P(0^+)$ is much smaller than $P(0^+)\!\approx\! 0.18$ resulting from the Edwards flat measure hypothesis.
\begin{figure}
\center{ 
\resizebox{0.47\textwidth}{!}{
\includegraphics{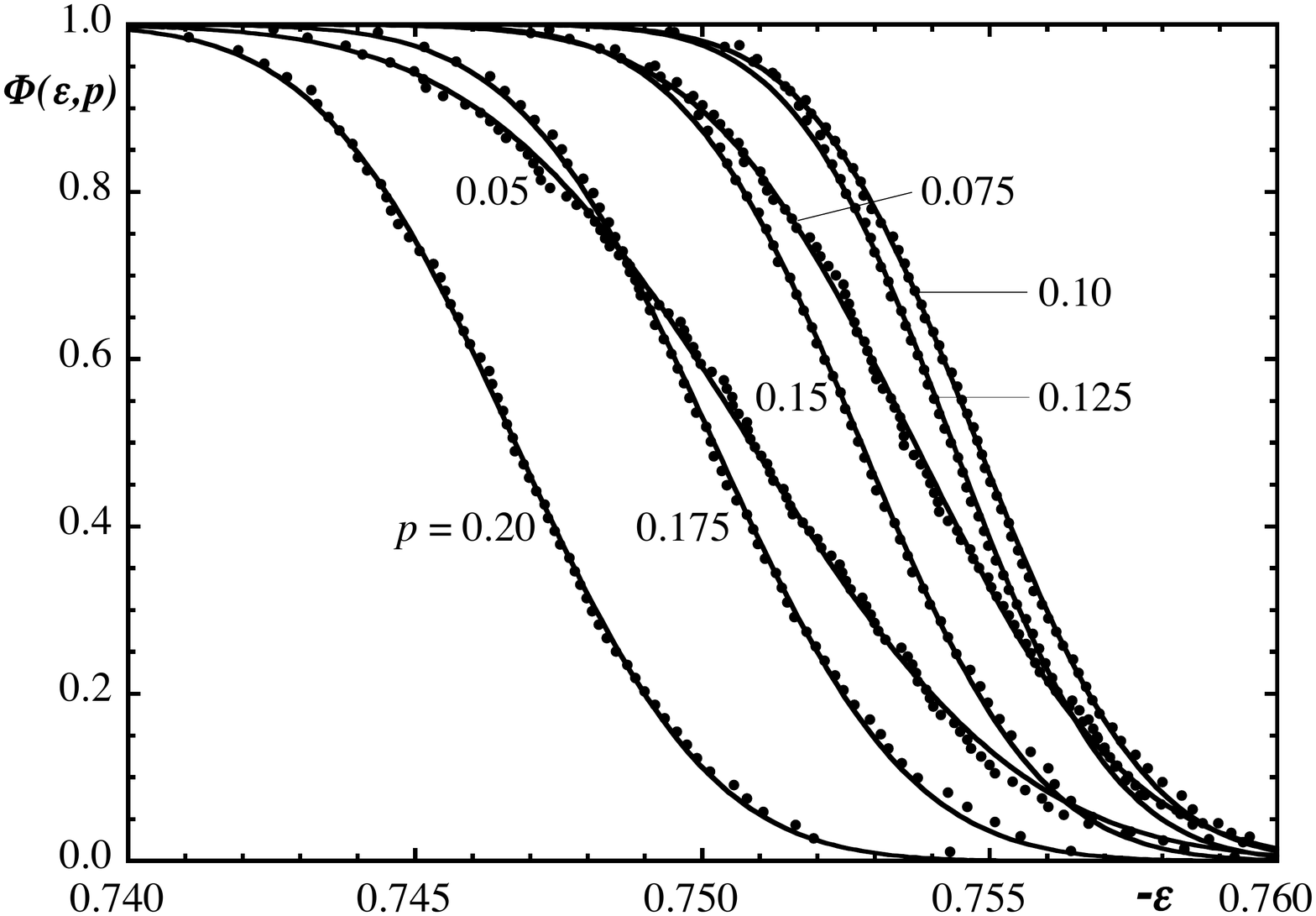}}}%
\caption{Integrated distribution of energies $\Phi(\epsilon)$ for different tapping strengths $p$ using random tapping wit $N\!=\!2500$ and $n=1000$.}
\label{P(E,tap)} 
\end{figure}

As a second test of the validity of the Edwards flat measure hypothesis the distribution of energies of meta\-stable states $P_{MS}(\epsilon)$ for different tapping strengths are compared \cite{DL03}. 

Counting how often an energy $E_l\!=\!N\epsilon_l<N\epsilon$ is found within $L$ Monte Carlo runs, the integrated distribution of energies of metastable states
\be
\Phi(\epsilon) =\frac{1}{L}\sum_l \Theta(\epsilon-\epsilon_l)
=\int_{\epsilon}^{\infty}\!\d\epsilon'\,P_{MS}(\epsilon')
\ee
results. This quantity is shown in Fig.\,\ref{P(E,tap)} obtained from random tapping with strength $p$ on systems of size \mbox{$N\!=\!2500$}.

Adopting a Gaussian distribution of energies of meta-stable states, 
Eq.~\req{beap},the integrated distribution is
\bel{bsiz}
\Phi(\epsilon)=\half \mbox{erfc}\Big(\sqrt{\frac{N}{2}}\,\frac{\epsilon-\bar\epsilon}{\delta}\Big).
\ee
Combinig this with Eq.~\req{bsaq}, which is a consequence of the Edwards hypothesis, the only effect of tapping should be a shift of the peak energy 
$\bar\epsilon$, whereas $\delta$ should be constant. The data in Tab.\,\ref{Fit} show that this is not the case.

The effective temperature $1/\beta_{tap}$ in Eq.~\req{bsaq} is supposed to depend on the tapping strength $p$. More precisely 
\bel{kzty}
\bar\epsilon-\delta^2\beta_{tap}=const.
\ee 
is expected. The existence of a finite optimal 
tapping \linebreak
strength implies that this dependence can not be 
mono\-tonous. For thermal tapping similar distributions are found. Again $\delta$ is not constant and Eq.~\req{kzty} is not fulfilled assuming $\beta_{tap}=1/ T_{tap}$.

\begin{table}
\center{
\begin{tabular}{ccc}
\hline\noalign{\smallskip}
p \% & $-\bar\epsilon$ &  $\delta$  \\
\noalign{\smallskip}\hline\noalign{\smallskip}
5.0 & 0.7509 & 0.263 \\ 
7.5 & 0.7537 & 0.207 \\
10.0 & 0.7548 & 0.163 \\
12.5 & 0.7544 & 0.157 \\
15.0 & 0.7526 & 0.172 \\
17.5 & 0.7504 & 0.188 \\
20.0 & 0.7468 & 0.189 \\
\noalign{\smallskip}\hline
\end{tabular} \ \ 
\begin{tabular}{ccc}
\hline\noalign{\smallskip}
$T_{tap}$ & $-\bar\epsilon$ &  $\delta$  \\
\noalign{\smallskip}\hline\noalign{\smallskip}
0.4 & 0.7458 & 0.227 \\ 
0.5 & 0.7495 & 0.174 \\
0.6 & 0.7517 & 0.151 \\
0.7 & 0.7532 & 0.138 \\
0.8 & 0.7536 & 0.127 \\
0.9 & 0.7532 & 0.128 \\
1.0 & 0.7512 & 0.134 \\
\noalign{\smallskip}\hline
\end{tabular}}
\caption{$\bar\epsilon$ and $\delta$ obtained from a fit of
$\Phi(/\epsilon)$ to Eq.\,\req{bsiz} for random and thermal tapping,
respectively.}
\label{Fit} 
\end{table}
%
\section*{Appendix A: Flat average over metastable\linebreak states}
Performing a flat or biased average over all metastable states, the local field distribution, say at site $o$ is
\bear
&&P(k)=Z^{-1}\Tr_{\sigma_o} \prod_{i (\ne o)}\!\!
\Tr_{\sigma_i}\overline{\;\delta(k-k_o) \e^{\beta k_o / 2}\;}
\nonumber\\ 
&&\qquad \; \overline{\times\!\prod_{i (\ne o)}
\Theta (k_i)\e^{\beta k_i / 2}}^J
\Theta(k)
\ear
with $k_o$ and $k_i$ given in \req{nfeq}. The metastable states are weighted with a Boltzmann factor $\e^{-\beta E}$ where \linebreak
$E\!=\!-\frac12\sum_i k_i$.
The effective temperature $1/\beta$ is assumed to characterize the tapping procedure. It might as well be viewed as Lagrange multiplier selecting the total energy of the metastable states under consideration.

The local field at a site $i\ne0$ is writen as $k_i\!=\!k'_i+\kappa_i$ with $\kappa_i\!=\!\sigma_o J_{oi} \sigma_i$. $k'_i$ is the local field of the system without the spin $\sigma_o$. 
For the SK-model in the limit $N\to\infty$
\bear
\!\!&&\overline{\prod_{i (\ne o)}
\Theta(k'_i+\kappa_i)}^J
\approx \prod_{i (\ne o)}\int_{-\kappa_i}^{\infty}\!\!\!\!\d k_i\,P(k_i)
\\
&&\qquad \approx \prod_{i (\ne o)}\Big\{1+\Theta(-\kappa_i)
\Big(\kappa_i P(0^+)-\half\kappa_i^2 P'(0^+)\Big)\Big\}.
\nonumber
\ear
Using the Fourier representation of the $\delta$-function and performing the average over $\kappa_i\!=\!\pm\frac{1}{\sqrt{N}}$
\bearl{evlz}
P(k)&=&Z^{-1}\!\!\int_{-i\infty}^{i\infty}\frac{\d\hat k}{2\pi}\,
\e^{(\hat k+\beta/2) k}
\nonumber\\
&\times&\Big\{1+\sfrac{1}{2N}
\Big(\hat k^2- \hat k P(0^+)-\half P'(0^+)\Big)\Big\}^N \Theta(k)
\nonumber\\
&=& \sqrt{\sfrac2\pi}\,\frac{\e^{-\frac12 [k-\frac12\{P(0^+)+\beta\}]^2}}
{1+{\rm erf}(\sfrac{1}{2\sqrt{2}}\{P(0^+)+\beta\})}\,\Theta(k)
\nonumber\\
&=& \sqrt{\sfrac2\pi}\,\frac{\e^{-\frac12 (k-\kappa)^2}}
{1+{\rm erf}(\sfrac{1}{\sqrt{2}}\kappa)}\,\Theta(k)
\ear
with $\kappa\!=\!\frac12\{P(0^+)+\beta\}$.
The energy $\epsilon$ per site is 
\bel{ljtx}
\epsilon=-\half\!\int\!\d k\,k\,P(k)=-\half\Big( P(0^+)+\kappa\Big).
\ee
%
\section*{Appendix B: Equation of motion for the two point function and drift velocity}
\label{App-B}
Let $R_{\sigma \sigma'}^{ij}(k,k')$ be the contribution  of sites $i$ and $j$ to \req{kury}. Flipping $\sigma_i\to -\sigma_i$ yields
\bear
\Delta_iR_{\sigma \sigma'}^{ij}(k,k')&=&-r(-k)R_{-\sigma \sigma'}^{ij}(-k,k')
-r(k)R_{\sigma \sigma'}^{ij}(k,k')
\nonumber\\
&-&2r(-k)P_{-\sigma}(-k)\partial_{k'}P_{\sigma'}(k').
\ear
The first two terms are due to $-k_i\to k_i$. The last term has its origin in $k_j\to k_j-2\sigma_iJ_{ij}\sigma_j$. Actually this term contains
$\overline{\av{\delta_{-\sigma,\sigma_i}\delta(k+k_i)
\delta_{\sigma',\sigma_j}\delta(k'-k_j)}}^{J'}$ with $J_{ij}\!=\!0$. This expression factorizes, however, for the SK-model. A corresponding contribution arises from flipping spin $\sigma_j$.

Flipping a spin on site $l\ne i,j$ yields diffusion terms and drift terms in analogy to \req{thjs}. The latter actually involve three point functions. In \cite{Ho90} a closure has been proposed on the basis of a generalized Kirkwood superposition approximation. A slight simplification yields
\bear
\sum_l\Delta_lR_{\sigma \sigma'}^{ij}(k,k')&\approx&
-\bigg\{\partial_k \Big( v_\sigma(k)-D(t)\partial_k \Big)
\nonumber\\
&& \;\;\; +[k\to k']\bigg\} R_{\sigma \sigma'}^{ij}(k,k').\qquad 
\ear
Collecting the above contributions
\bearl{ktsq}
\partial_t R_{\sigma \sigma'}(k,k')&\approx&
-r(-k)R_{-\sigma \sigma'}(-k,k')-r(k)R_{\sigma \sigma'}(k,k')
\nonumber\\
&-&\partial_k \Big( v_\sigma(k)-D\partial_k\Big)
R_{\sigma \sigma'}(k,k')
\nonumber\\
&-&2r(-k)P_{-\sigma}(-k)\partial_{k'}P_{\sigma'}(k')
\nonumber\\
&+& \Big[\sigma\to \sigma', \ k\to k'\Big].
\ear
This yields the equation of motion for the drift velocity \req{ftho}.
There are different types of contributions. The first two lines of \req{ktsq} result in corresponding expressions for $v_\sigma(k)P_\sigma(k)$. The contribution of the third line is easily taken into account, because it factorizes. The terms analogous to the first two lines acting on $\sigma'$ and $k'$ can not be expressed in terms of $P$ and $v$ only. Taking into account 
the identities \req{thjs} and \req{szhw} they are approximated by adding contributions $\sim v_\sigma(k)P_\sigma(k)$ and $\sim k P_\sigma(k)$. 
The resulting equation of motion for $v_\sigma(k)$ is
\bearl{hatb}
\partial_t v_\sigma(k)&=&-r(-k)\Big(v_\sigma(k)+v_{-\sigma}(-k)+4\avav{r'\big.}\!\Big)\frac{P_{-\sigma}(-k)}{P_{\sigma}(k)}
\nonumber\\
&-&\Big( v_\sigma(k)-D\partial_k-2 D \Big[\partial_k\ln(P_\sigma(k))\Big]
\Big) \partial_k v_{\sigma}(k)
\nonumber\\
&+&4\av{r\bar r}\Big[\partial_k \ln(P_\sigma(k))\Big] +A\, v_\sigma(k)+B\,k.
\ear
For $k\to\infty$ the local field distribution is expected to follow $P_\sigma(k\gg1)\sim \e^{-\frac12k^2}$. This is the case if \req{clis} holds not only for the initial state, but for all time in the limit $k$ or $k'\to\infty$. The coefficients $A$ and $B$ are then determined such that this asymptotic behavior and the sum rule \req{nsza} is fulfilled.
%
\begin{acknowledgement} 
I like to thank Mike Moore for discussions and Paul Eastham for helpful correspondence, especially on the Edwards hypothesis and the resulting local field distribution.
\end{acknowledgement}


\end{document}